\documentclass[letterpaper, 10 pt, conference]{ieeeconf}  

\IEEEoverridecommandlockouts                              
\pdfoutput=1
\usepackage{graphics} 
\usepackage{epsfig} 
\usepackage{mathptmx} 
\usepackage{times} 
\usepackage{amsmath} 
\usepackage{amssymb}  
\usepackage[ruled]{algorithm2e}
\usepackage{bm}
\usepackage{stfloats}
\usepackage{subfigure}
\usepackage{graphicx}
\usepackage{threeparttable}

\title{\LARGE \bf
Decentralized Coordination Between Economic Dispatch and Demand Response in Multi-Energy Systems
}

\author{Zishun Liu, Shanying Zhu, Jinming Xu, Cailian Chen 
\thanks{This work was supported by NSF of China under the grants 62003302, 61922058, 62025303, and 61933009.}
\thanks{Zishun Liu, Shanying Zhu and Cailian Chen is with the Department of Automation, Shanghai Jiaotong University, and also with Key Laboratory of System Control and Information Processing,
	Ministry of Education of China, Shanghai 200240, China, and also with Shanghai Engineering Research Center of Intelligent Control
	and Management, Shanghai 200240, China.
        {\tt\small liu0105@sjtu.edu.cn}, {\tt\small shyzhu@sjtu.edu.cn}{}}%
\thanks{Jinming Xu is with College of Control Science and Engineering, Zhejiang University, Hangzhou 310027, Zhejiang, China.%
}
}

\begin{document}

\maketitle
\thispagestyle{empty}
\pagestyle{empty}

\begin{abstract}

In this paper, we investigate the problem of coordination between economic dispatch (ED) and demand response (DR) in multi-energy systems (MESs), aiming to improve the economic utility and reduce the waste of energy in MESs. Since multiple energy sources are coupled through energy hubs (EHs), the supply-demand constraints are nonconvex. To deal with this issue, we propose a linearization method to transform the coordination problem to a convex social welfare optimization one. Then a  decentralized algorithm based on parallel Alternating Direction Method of Multipliers (ADMM) and dynamic average tracking protocol is developed, where each agent could only make decisions based on information from their neighbors. Moreover, by using variational inequality and Lyapunov-based techniques, we show that our algorithm could always converge to the global optimal solution. Finally, a case study on the modified IEEE 14-bus network verifies the feasibility and effectiveness of our algorithm.

\end{abstract}

\section{INTRODUCTION}

Recent developments on technologies of control, optimization and communication enable multi-energy systems (MESs) to coordinate multiple energy sources, e.g., electricity, gas and solar, and form complementary advantages during the process of production, transformation, transaction and consumption \cite{2020Multienergy}. Compared to power systems, MESs can satisfy the demand of energy users more precisely and reduce redundant conversion between different energy forms, thus promoting energy utility. 

How to manage energy is one of the most important issue in MESs, especially the problem of economic dispatch (ED) on the supply side and demand response (DR) on the demand side. Extending from power systems, ED problem in MESs is a resource allocation problem essentially which aims at minimizing the generation cost by allocating energy source properly \cite{2008Decentralized}, and DR problem in MESs seeks to maximize benefits at the demand side by making informed decisions with respect to the energy consumption \cite{2008A}. Unlike traditional power systems, the coupling among different kinds of energy makes it more difficult in problem formulation. To describe the coupling mathematically, Geidl \textit{et al}. proposed the concept of energy hub (EH) in \cite{4077107}, which has been widely adopted in the literature. 
Based on EHs, optimization methods like mathematical programming methods, AI-based methods and hybrid methods have been used to solve the ED and DR problems in MESs, see the survey \cite{2017A} and reference therein. However, most of the algorithms are in a centralized form, which means they need a center to collect and handle global information. In large scale MESs, centralized algorithms will cost much on long-distance communication, and the center will bear very high risk since it stores global information of the whole systems. 

In recent years, many decentralized algorithms have been established and applied to energy systems, which only require each node to share information with its neighbors, and thus guarantee the security and privacy \cite{2019Network}. Most of the decentralized algorithms can be categorized into two types: one is consensus-based algorithm, which allows nodes to obtain global optimum by making local decisions and sharing information with neighboring nodes, see \cite{2014Consensus} and \cite{Chaojie2017Distributed}. Another type seeks to deal with its dual problem, which is often separable in the dual domain, see \cite{2017AS} and \cite{2018A}. It should be noted that these algorithms could only be applied to electrical power systems. As for the MESs,  multiple energy sources are coupled through EHs, making the supply-demand constraints nonconvex. Only a few works have been reported to deal with the coupling and nonlinearity in MESs. For example, \cite{2020wzb} proposes a decentralized algorithm based on sigmoid-like linearization operator and dual-based methods to solve ED problem in MES. Ref. \cite{2018Whole} decomposes the inner structure of EHs to deal with the nonconvexity and proposes an algorithm based on consensus-based methods to minimize the generation cost. However, these works only deal with ED or DR problem separately. 


Aiming at optimizing total welfare of both supply and demand sides, the coordination problem between ED and DR has attracted much attention. In \cite{2014Incremental}, the authors propose a decentralized incremental welfare consensus algorithm to solve the coordination problem by exchanging information locally. However, the objective functions should be strictly convex/concave. The work \cite{2015Distributed} extends \cite{2014Incremental} and proposes a decentralized algorithm by integrating the average consensus protocol and ADMM, but it needs extra steps to handle the energy mismatch between supply side and demand side. The work \cite{8356102} considers energy storage in the coordination problem and provides a consensus-based decentralized algorithm. Ref. \cite{2019Distributed} proposes a decentralized algorithm based on push-sum and gradient method to solve the optimal energy resource condition over time-varying directed communication networks. Moreover, an accelerated distributed algorithm is proposed to improve the convergence speed and reduce the communication burden. One main drawback of \cite{8356102} and \cite{2019Distributed} is that their objective functions are restricted to quadratic forms, and the algorithms are only applicable to electrical power systems.

Motivated by the above observations, in this paper we solve the coordination problem between ED and DR by employing parallel-ADMM and dynamic average tracking technique in MESs. The contributions of this paper are as follows:

\textit{a$\left.\right)$} The coordination problem between ED and DR in MESs is formulated as a social welfare optimization problem (SWOP). Since multiple energy sources are coupled through EHs, the supply-demand constraints are nonconvex. This contrast to the convex formulation in electrical power systems in \cite{2014Incremental}$\sim$\cite{2019Distributed}. By applying dimension augmentation techniques, we linearize the nonlinear constraints and uncouple the variables to transform the SWOP as a convex optimization problem. 

\textit{b$\left.\right)$} A decentralized algorithm based on parallel-ADMM and dynamic average tracking protocol is proposed, which only requires each node to make local decisions and share information with its neighbors. Different from \cite{2014Incremental,8356102,2019Distributed}, our algorithm does not assume the cost functions to be strictly convex or quadratic. By using variational inequality and Lyapunov-based techniques, it is shown that our algorithm can solve the SWOP for all convex objective functions.

The paper is organized as follows. In Section \ref{s2} we formulate the coordination problem between ED and DR as a SWOP. In Section \ref{s3}, we transform the original SWOP to a convex optimization problem and propose a decentralized algorithm based on parallel-ADMM and dynamic average tracking protocol. Convergence analysis of our algorithm is provided in Section \ref{s4}. Section \ref{s5} presents a case study to illustrate effectiveness of the proposed algorithm. Finally, Section \ref{s6} concludes the paper.

\begin{figure}[t]	
	\centering
	\includegraphics[width=0.35\textwidth]{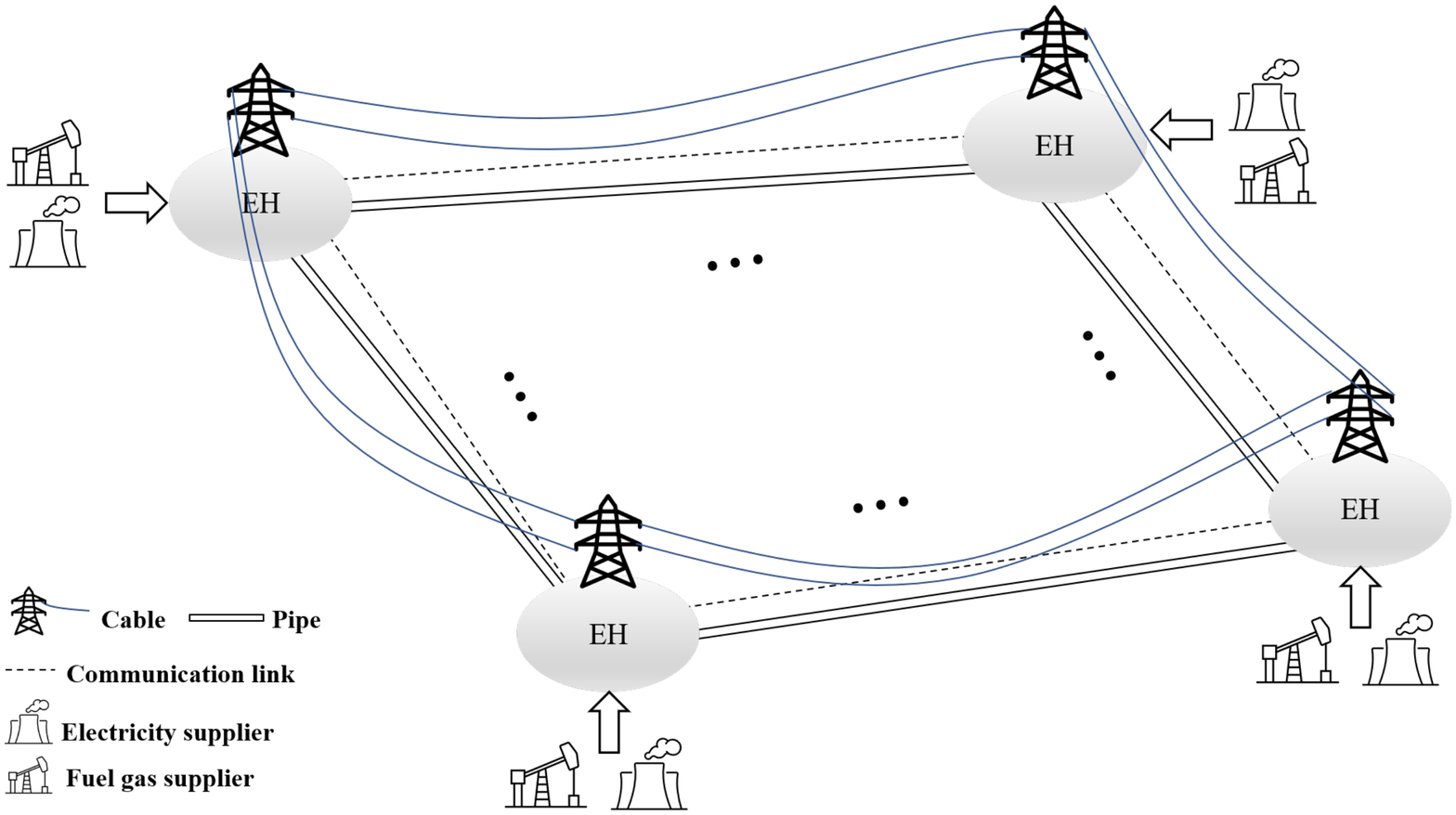}
	\caption{Structure of MES.}
	\label{fig-1}
\end{figure}

\textit{Notation}. For a vector $x$, $x_i^k$ denotes that it belongs to $\text{EH}_i$ in $k$th iteration, $x^T$ denotes its transpose,
 $\left\|x\right\|_2$ denotes its Euclidean norm. For a series of column vector $x_1,x_2,\dots,x_n$ in the same length, vector $\bm{x}$ denotes $[x_1^T,x_2^T,\dots,x_n^T]^T$.  
For a matrix $P$, $P=[p_{ij}]_{n}$ denotes a $n\times n$ matrix, $I_n$ denotes $n\times n$ identity matrix, $\mathcal{O}_n$ denotes a $n\times n$ matrix with all entries equal to 1/$n$, $P\succ 0$ denotes that it is positive definite, $P^T$ denotes its transpose, $\left\|P\right\|$ denotes its Euclidean norm, $\left\|x\right\|_P=x^TPx$, $\Gamma(P)$ denotes its spectral radius. 
The Kronecker product is denoted by $\otimes$.




\begin{figure}[t]	
	\centering
	\includegraphics[width=0.35\textwidth]{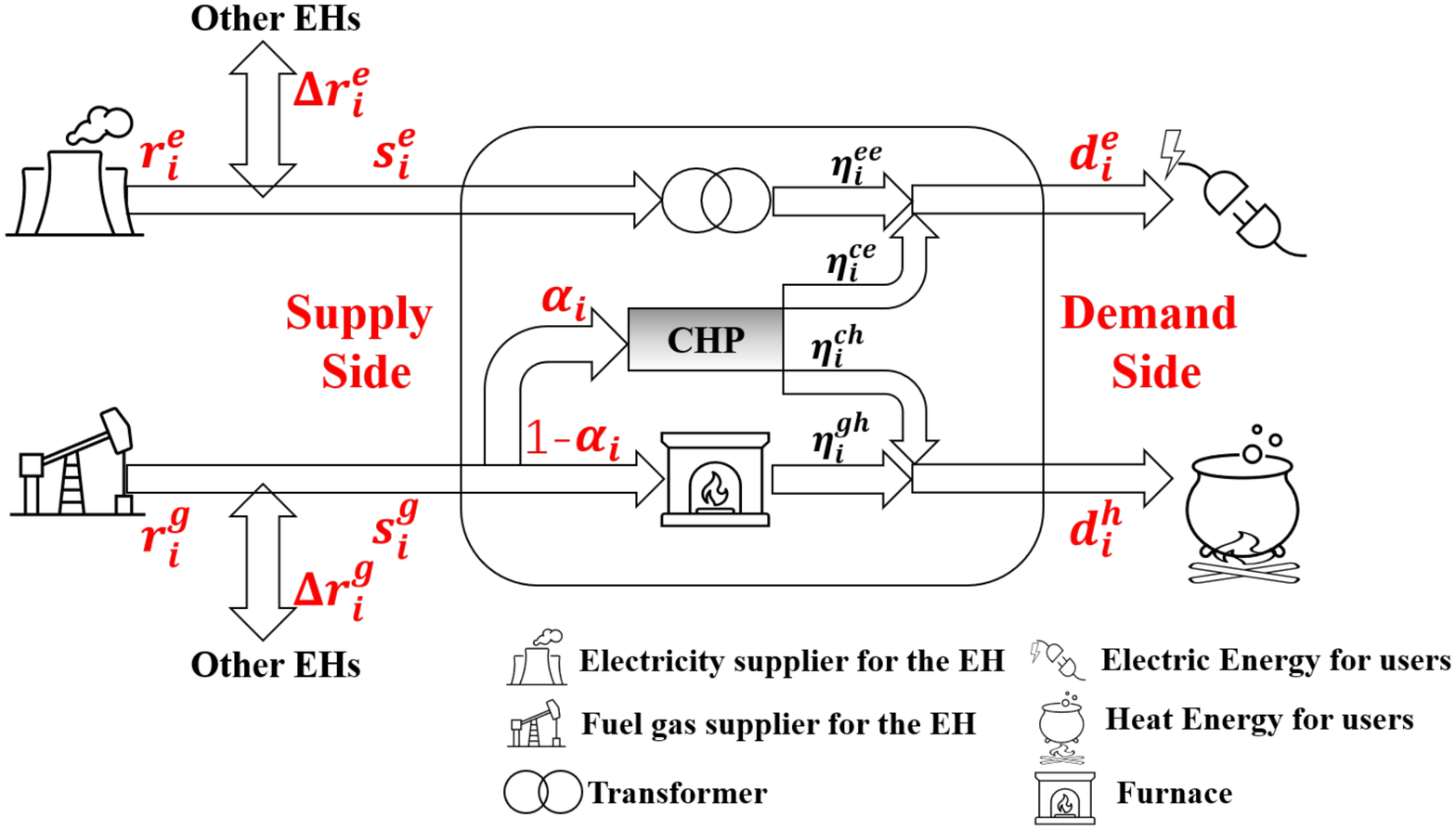}
	\caption{Structure of EH}
	\label{fig-2}
\end{figure}

\section{PROBLEM FORMULATION}\label{s2}

In this part, we will introduce the structure of MESs considered in this paper, and formulate the coordination problem between ED and DR in MESs.

\subsection{System Model}\label{2a}

An MES is consisted of $n$ EHs, channels between EHs, pipes and cables for energy transfer, energy suppliers on the supply side and energy users on the demand side. Fig.\ref{fig-1} shows the basic structure of the MES.

The function of EHs is to coordinate the energy at a local part inside an MES. They get energy source from energy suppliers and transform it to the energy required by the demand side, as well as trade on energy source with other EHs. In this paper, based on typical cases, we assume that the EH is a two-input and two-output system. It transforms high-voltage electricity and fuel gas to electricity and heat power available for the demand side via a transformer, a CHP and a furnace. Fig.\ref{fig-2} shows the structure and function of the EH, in which $s_i=[s_i^e,s_i^g]^T$ is the energy source that $\text{EH}_i$ needs on its supply side, $d_i=[d_i^e,d_i^g]^T$ is the energy offered to the demand side of $\text{EH}_i$, $\eta_i^{ee}$, $\eta_i^{ce}$, $\eta_i^{ch}$ and $\eta_i^{gh}$are the efficiency coefficients of transformation, and $\alpha_i\in[0,1]$ is the percentage of fuel gas supplied to the CHP, which is an optimization variable. Based on Fig.\ref{fig-2}, we can give the coupling matrix $A_i$:
$$A_i=\left[\begin{matrix}
	\eta_i^{ee} & \alpha_i\cdot\eta_i^{ce} \\
	0 & \alpha_i\cdot\eta_i^{ch}+(1-\alpha_i)\cdot\eta_i^{gh}\\
\end{matrix}\right]_.$$
Then the relationship between the supply side and the demand side can be established by
 \begin{equation}\label{e8}
 	d_i=A_is_i.
 \end{equation}

In the process of energy supplying, $r_i=[r_i^e,r_i^g]^T$ is the amount of energy source that electricity and fuel gas suppliers supply for $\text{EH}_i$. The cost function of generating and producing fuel gas is $C_i(r_i)=Ce_i(r_i^e)+Cg_i(r_i^g)$. It should be pointed out that usually $C_i(r_i)$ is in quadratic form \cite{8356102}. $\Delta r_i=[\Delta r_i^e,\Delta r_i^g]^T$ is the amount of energy that $\text{EH}_i$ trades with other EHs. Obviously, we have $s_i=r_i+\Delta r_i$. At the system level, the transaction must obey the following conservation property:
\begin{equation}\label{e7}
	\sum\limits_{i=1}^n{s_i}=\sum\limits_{i=1}^n{r_i}.
\end{equation}

Additionally, $d_i$ may not precisely satisfy the demand side. In this model, we use $U_i(d_i)=E_i(d_i)-T(d_i)$ to describe the total utility of the demand side of $\text{EH}_i$, in which $E_i(d_i)$ is the total earning contributed by $d_i$, and $T(d_i)=\theta_i\cdot\left\| d_i-\hat{d_i} \right\|_2^2$ is the Taguchi loss function \cite{Taguchi2007Taguchi} to measure the degree of dissatisfaction, $\theta_i$ is a parameter decided by the system, and $\hat{d_i}$ is the expected energy consumption of the consumers on the demand side of $\text{EH}_i$.

Considering the connection among EHs mentioned above, the MES can be modeled as a connected and undirected graph $\mathcal{G}=\{\mathcal{N},\mathcal{E}\}$, whose nodes in $\mathcal{N}$ are EHs which could communicate, trade and transform energy source through edges in $\mathcal{E}$. Define $\mathcal{W}=\left[w_{ij}\right]_n$ as the weight matrix of $\mathcal{G}$, based on \cite{2004Fast}, $\mathcal{W}$ should satisfy the following conditions:
\begin{enumerate}
	\item\label{con1} For $\forall i,j=1,2,\dots,n$, $w_{ij}\in [0,1]$, and $w_{ij}=0$ means $\text{EH}_i$ and $\text{EH}_i$ are not directly connected by channels.
	\item\label{con2} $\mathcal{W}$ is symmetric.
	\item\label{con3} $\mathcal{W}$ is doubly stochastic, which means:
	\begin{equation}\label{e4}
			\sum\limits_{i=1}^{n}{w_{ij}}=1,\quad\sum\limits_{j=1}^{n}{w_{ij}}=1,\quad\forall i,j=1,2,\dots n.
	\end{equation}
\end{enumerate}

\subsection{Formulation of SWOP}

For each $\text{EH}_i$, we use utility at the demand side minus cost at the supply side to describe the local welfare. Define local welfare function $WE_i$ as follows 
\begin{equation}\label{e9}
	WE_i=U_i(d_i)-C_i(r_i)-\zeta^T\Delta r_i,
\end{equation}
in which $\zeta=[\zeta^e,\zeta^g]^T$ is the price of energy source in trade. Summing (\ref{e9}) up,  we can get our problem as follows
\begin{equation}\label{e10}
	\max\limits_{\bm{r,s,d,\alpha}}\sum\limits_{i=1}^n{WE_i}.
\end{equation}

Applying (\ref{e7}) into (\ref{e10}), we can get $\sum_{i=1}^n\Delta r_i=0$. Then we can formulate our problem as an optimization problem constrained by (\ref{e8}), (\ref{e7}), and generation and load capacities of each $\text{EH}_i$:
\begin{subequations}\label{e12}
	\begin{align}
		\min\limits_{\bm{r},\bm{s},\bm{d},\bm{\alpha}}F&=\sum\limits_{i=1}^n{[C_i(r_i)-U_i(d_i)]}\\
		\text{s.t}\quad\sum\limits_{i=1}^n{s_i}&=\sum\limits_{i=1}^n{r_i}, \\
		d_i&=A_is_i,\\
		\underline{r_i}\leqslant &r_i\leqslant\overline{r_i},\\
		\underline{s_i}\leqslant &s_i\leqslant\overline{s_i},\\
		\underline{d_i}\leqslant &d_i\leqslant\overline{d_i},\\
		0\leqslant &\alpha_i\leqslant1, \quad\forall i=1,2,\dots, n.
	\end{align}
\end{subequations} 
where $\underline{r_i},\underline{s_i},\underline{d_i}$ and $\overline{r_i},\overline{s_i},\overline{d_i}$ are the lower and upper limits of the variables, respectively.

It should be pointed out that the problem above is nonconvex since optimization variables $s_i$, $d_i$ and $\alpha_i$, $\forall i=1,2,\dots n$ are coupled with each other in the constraint (\ref{e12}c), making it a nonlinear equality. Several centralized methods have been proposed in the existing literature to solve such nonconvex problem \cite{2017A}. Our goal in this paper is to design a fully decentralized algorithm to solve the above non-convex problem only based on local information.


\section{A DECENTRALIZED COORDINATION ALGORITHM BASED ON PARALLEL-ADMM}\label{s3}
In this section, we will develop a decentralized coordination algorithm based on parallel-ADMM and dynamic average tracking.
\subsection{Convexification of the Problem}\label{s3a}

To deal with the nonlinearity in problem (\ref{e12}), we introduce the concept of hypothetical ports as in \cite{2020Multienergy} into our problem. By applying this concept, we can split the fuel gas input port of $\text{EH}_i$ into two ports which supply fuel gas at certain percentage: one supplies $\alpha_is_i^g$ and the other supplies $(1-\alpha_i)s_i^g$. In this way the energy supply stays the same with the original SWOP.

Based on the discussion above, define
$$l_i=[l_{i1},l_{i2},l_{i3}]^T=[s_i^e,\alpha_is_i^g,(1-\alpha_i)s_i^g]^T,$$ 
$$B_i=\left[\begin{matrix}
	\eta_i^{ee} & \eta_i^{ce} & 0\\
	0 & \eta_i^{ch} & \eta_i^{gh}\\
\end{matrix}\right],\quad \forall i=1,2,\dots, n.$$

Noticing that $B_i$ is independent of the optimization variables, the constraint (\ref{e12}c)  can be rewritten as a linear constraint:
\begin{equation}\label{e13}
	B_il_i=d_i.
\end{equation}
Meanwhile, the constraint (\ref{e12}g) is equivalently expressed as 
\begin{equation}\label{eee1}
	l_{i2}\geqslant0,\quad l_{i3}\geqslant0.
\end{equation}

Define
$$ u_i=[l_i^T,d_i^T]^T,\quad \bar{B_i}=B_i\left[\begin{matrix}
	1&0&0&0&0\\
	0&1&0&0&0\\
	0&0&1&0&0\\
\end{matrix}\right],$$
$$M=\left[\begin{matrix}
	1&0&0&0&0\\
	0&1&1&0&0\\
\end{matrix}\right],\quad
M_1=\left[\begin{matrix}
	0&0&0&1&0\\
	0&0&0&0&1\\
\end{matrix}\right],$$
$$M_2=\left[\begin{matrix}
	0&1&0&0&0\\
\end{matrix}\right],\quad
M_3=\left[\begin{matrix}
	0&0&1&0&0\\
\end{matrix}\right],
$$
then we have $B_il_i=\bar{B_i}u_i$, $s_i=Mu_i$, $d_i=M_1u_i$.
Based on these relations, the constraint (\ref{e12}b), (\ref{e12}c) (\ref{e12}e), (\ref{e12}f), (\ref{e13}) and (\ref{eee1}) can be rewritten as
\begin{subequations}\label{e15}
	\begin{align}
		\sum\limits_{i=1}^n{r_i}&=\sum\limits_{i=1}^n{Mu_i},\\
		\bar{B_i}u_i&=M_1u_i,\\
		\underline{s_i}\leqslant &Mu_i\leqslant\overline{s_i},\\
		\underline{d_i}\leqslant &M_1u_i\leqslant\overline{d_i},\\
		&M_2u_i\geqslant0,\\
		&M_3u_i\geqslant0.
	\end{align}
\end{subequations}

 We introduce the indicator function as follows
\begin{equation}\label{e16}
	I(r_i,u_i,\Omega_i)=\begin{cases}
		0,\quad\quad if\quad (r_i,u_i)\in\Omega_i,\\
		+\infty,\quad otherwise,
	\end{cases}
\end{equation} 
where $\Omega_i=\left\{(r_i,u_i)|\text{(\ref{e12}d), (\ref{e15}b)}\sim\text{(\ref{e15}f),} \right\}$
is a convex set.

Then, the problem (\ref{e12}) can be transformed to the following convex problem
\begin{subequations}\label{e17}
	\begin{align}
		\min\limits_{\bm{r,u}}F=\sum\limits_{i=1}^n{F_i(r_i,u_i)}=&\sum\limits_{i=1}^n{[C_i(r_i)-U_i(M_1u_i)+I(r_i,u_i,\Omega_i)]}\\
		\text{s.t.}\quad \sum\limits_{i=1}^n{r_i}&=\sum\limits_{i=1}^n{Mu_i},\quad i=1,2,\dots,n.
	\end{align}
\end{subequations}

\subsection{Decentralized Algorithm Design}

In this subsection, we will develop a decentralized algorithm to solve problem (\ref{e17}). However, we can see that all the optimization variables are coupled in one global constraint (\ref{e17}b). To uncouple them, we introduce two groups of additional variables $x_1,x_2,\dots,x_n$ and $y_1,y_2,\dots,y_n$ into our problem. 
Define $$\Pi=\bigl\{(\bm{x},\bm{y})\text{ }|\text{ }\sum\limits_{i=1}^n{x_i}=\sum\limits_{i=1}^n{y_i},\text{ }i=1,2,\dots, n\bigr\},$$ then problem (\ref{e17}) can be reformulated as
\begin{subequations}\label{e19}
	\begin{align}
		\min\limits_{\bm{r},\bm{u},\bm{x},\bm{y}}J=&\sum\limits_{i=1}^n{[F_i(r_i,u_i)+I(\bm{x},\bm{y},\Pi)]}\\
		\text{s.t.}\quad &x_i=r_i,\quad y_i=Mu_i,\text{ }i=1,2,\dots,n.
	\end{align}
\end{subequations}

The augmented Lagrange function of (\ref{e19}) is
\begin{equation*}\label{e20}
	\begin{aligned}
		&{L_{\rho}}(\bm{r},\bm{u},\bm{x},\bm{y},\bm{p},\bm{q})=\sum\limits_{i=1}^n{[\text{ }F_i(r_i,u_i)+p_i^T(r_i-x_i)+q_i^T(Mu_i-y_i)}\\
			&{+\frac{\rho}{2}\left\|r_i-x_i\right\|_2^2+\frac{\rho}{2}\left\|Mu_i-y_i\right\|_2^2\text{ }]}\\
			&\overset{def}{=}\sum\limits_{i=1}^n{{L_i}_\rho(r_i,u_i,x_i,y_i,p_i,q_i)},\quad (\bm{x},\bm{y})\in\Pi,(r_i,u_i)\in\Omega_i,
	\end{aligned}
\end{equation*} 
in which ${L_i}_\rho$ is the local augmented Lagrange function of $\text{EH}_i$.
We can apply the ADMM \cite{Glowinski1975} to solve the problem (\ref{e19}) as follows
\begin{subequations}\label{e21}
	\begin{align}
		\left[\begin{matrix}
			r_i^{k+1}\\
			u_i^{k+1}\\
		\end{matrix}\right]&=\arg\min\limits_{(r_i,u_i)\in\Omega_i}{L_i}_\rho(r_i,u_i,x_i^k,y_i^k,p_i^k,q_i^k), \\
		\left[\begin{matrix}
			\bm{x}^{k+1}\\
			\bm{y}^{k+1}\\
		\end{matrix}\right]&=\arg\min\limits_{(\bm{x},\bm{y})\in\Pi}L_\rho(\bm{r}^{k+1},\bm{u}^{k+1},\bm{x},\bm{y},\bm{p}^k,\bm{q}^k), \\
		p_i^{k+1}&=p_i^k+\rho(r_i^{k+1}-x_i^{k+1}),\\
		q_i^{k+1}&=q_i^k+\rho(Mu_i^{k+1}-y_i^{k+1}).
	\end{align}
\end{subequations}

Following the parallel ADMM procedure proposed in \cite{NP1995}, we can get that $\bm{p}^k=-\bm{q}^k\overset{def}{=}\bm{\mu}^k$, and steps in (\ref{e21}) can be simplified as follows
\begin{subequations}\label{eee2}
	\begin{align}
		\left[\begin{matrix}
			r_i^{k+1}\\
			u_i^{k+1}\\
		\end{matrix}\right]&=\arg\min\limits_{(r_i,u_i)\in\Omega_i}F_i(r_i,u_i)+(\mu^k)^T(r_i-Mu_i) \notag\\
	    &+\frac{\rho}{2}\left\|(r_i-r_i^k)+\frac{1}{2n}\sum\limits_{i=1}^n{(r_i^{k}-Mu_i^{k})}\right\|_2^2\notag\\
	    &+\frac{\rho}{2}\left\|\frac{1}{2n}\sum\limits_{i=1}^n{(r_i^{k}-Mu_i^{k})-M(u_i-u_i^k)}\right\|_2^2,\\
		\mu^{k+1}&=\mu^k+\frac{\rho}{2n}\sum\limits_{i=1}^n{(r_i^{k+1}-Mu_i^{k+1})}.
	\end{align}
\end{subequations}

It can be seen that the only part that requires global information is $\frac{1}{n}\sum\limits_{i=1}^n{(r_i^{k+1}-Mu_i^{k+1})}$. Our method to deal with this part is using dynamic average tracking technique \cite{2018A,7402509}. To be specific, we use $e_i^{k+1}=\sum\limits_{j=1}^n{w_{ij}e_j^k}+(r_i^{k+1}-r_i^k)-(Mu_i^{k+1}-Mu_i^k)$ to replace the centralized calculation of  the global quantity, and use $\mu_i^{k+1}=\sum\limits_{j=1}^n{w_{ij}\mu_j^k}+\rho e_i^{k+1}$ to calculate $\mu_i$ in decentralized form, in which $\sum\limits_{j=1}^n{w_{ij}e_j^k}$ and $\sum\limits_{j=1}^n{w_{ij}\mu_j^k}$ can be treated as average value of $e_i^k$ and $\mu_i^k$, respectively. Additionally, we merge two penalty items in (\ref{eee2}a) into one part. Define $u_i^k=[u_{i1}^k,\dots,u_{i5}^k]^T$, the proposed algorithm is formally organized in \textit{Algorithm~\ref{a1}}. 
\begin{algorithm}\label{a1}
	\caption{Decentralized Coordination Algorithm Based on Parallel-ADMM at Each $\text{EH}_i$}
	\LinesNumbered
	\KwIn{$r_i^0$, $u_i^0$, $\mu_i^0$, $e_i^0$, $\sigma_i^0$, $\phi_i^0$, $\epsilon$, $N$, $\mathcal{W}=\left[w_{ij}\right]_{n}$}
	\KwOut{Optimal solution $r_i^{*}$, $s_i^{*}$, $d_i^{*}$, $\alpha_i^{*}$}
	
	Initialization: $e_i^0=r_i^0-Mu_i^0$, $r_i^0$. $u_i^0$, $\mu_i^0$, $e_i^0$, $\sigma_i^0$, $\phi_i^0$ can be set randomly, $\epsilon$ should be set by a small value (less than $0.1e_i^0$), $N$ should be set by a big value ($>100$), $\mathcal{W}$ should meet the conditions in Section \ref{2a}.\\
	\While{$\left\|r_i^{k+1}-r_i^k\right\|\geqslant\epsilon$, or $\left\|s_i^{k+1}-s_i^k\right\|\geqslant\epsilon$, or $\left\|d_i^{k+1}-d_i^k\right\|\geqslant\epsilon$, or $\left\|\alpha_i^{k+1}-\alpha_i^k\right\|\geqslant\epsilon$ or $k<N$ }{
		$\sigma_i^k=\sum\limits_{j=1}^n{w_{ij}e_j^k}$; \\
		$\phi_i^k=\sum\limits_{j=1}^n{w_{ij}\mu_j^k}$; \\
		$\left[\begin{matrix}
			r_i^{k+1}\\
			u_i^{k+1}\\
		\end{matrix}\right]=\arg\min\limits_{(r_i,u_i)\in\Omega_i}[\text{ }F_i(r_i,u_i)+(\phi_i^k)^T(r_i-Mu_i)+\frac{\rho}{2}\left\|(r_i-r_i^k)-M(u_i-u_i^k)+\sigma_i^k\right\|_2^2\text{ }]$ ;\\
	    $e_i^{k+1}=\sigma_i^k+(r_i^{k+1}-r_i^k)-(Mu_i^{k+1}-Mu_i^k)$; \\
	    $\mu_i^{k+1}=\phi_i^k+\rho e_i^{k+1}$;\\
	    $s_i^{k+1}=\left[\begin{matrix}
	    	u_{i1}^{k+1},&u_{i2}^{k+1}+u_{i3}^{k+1}\\
	    \end{matrix}\right]^T$, $d_i^{k+1}=\left[\begin{matrix}
	    	u_{i4}^{k+1},&u_{i5}^{k+1}\\
	    \end{matrix}\right]^T$, $\alpha_i^{k+1}=\dfrac{u_{i2}^{k+1}}{u_{i2}^{k+1}+u_{i3}^{k+1}}$;\\
	    $k+1\to k$;\\
	}
\end{algorithm}

In \textit{Algorithm \ref{a1}}, it can be seen that the proposed algorithm is a fully decentralized one, which only requires local processing and local communication between neighboring nodes.

\section{Analysis of Convergence}\label{s4}

In this section, we will provide the convergence analysis of the proposed decentralized algorithm. To this end, we introduce two standard assumptions.

\textit{Assumption 1}\label{as1}: All the $C_i(r_i)$ are convex, and $U_i(d_i)$ are concave on $\Omega_i$, $\forall i=1,2,\dots ,n$.

\textit{Assumption 2}\label{as2}: Graph $\mathcal{G}=\{\mathcal{N},\mathcal{E}\}$ is connected and undirected.

Define $\bar{e}^k=\dfrac{1}{n}\sum\limits_{i=1}^ne_i^k$, $\bar{\mu}^k=\dfrac{1}{n}\sum\limits_{i=1}^n\mu_i^k$,  $\delta_i^k=(r_i^k-Mu_i^k)-\bar{e}^k$, $\Delta e_i^k=e_i^k-\bar{e}^k$, $\Delta\mu_i^k=\mu_i^k-\bar{\mu}^k$, $W=\mathcal{W}\otimes I_2$. Based on these definitions, we can prove the following two lemmas.

\textit{Lemma 1}: Under \textit{Assumption 2}, we have $\bar{e}^k=\dfrac{1}{n}\sum\limits_{i=1}^n\left(r_i^k-Mu_i^k\right)$, and $\bar{\mu}^{k+1}=\bar{\mu}^k+\rho\bar{e}^{k+1}$.

\textit{Proof}: Under \textit{Assumption 2}, using the condition (\ref{e4}), we can get that
\begin{equation}\label{e24}
	\sum\limits_{i=1}^n\sum\limits_{j=1}^n{w_{ij}e_j^k}=\sum\limits_{j=1}^n(\sum\limits_{i=1}^nw_{ij})e_j^k=\sum\limits_{j=1}^n{e_j^k}=\sum\limits_{i=1}^n{e_i^k}.
\end{equation}

Apply (\ref{e24}) to the equations in lines \textbf{3} and \textbf{6} in \textit{Algorithm~\ref{a1}}, we can get that
\begin{equation}\label{e25}
	\sum\limits_{i=1}^n{e_i^{k+1}}=\sum\limits_{i=1}^n{e_i^k}+\sum\limits_{i=1}^n{[(r_i^{k+1}-r_i^k)-(Mu_i^{k+1}-Mu_i^k)]}.
\end{equation}

Then using mathematical induction, we can easily get to the following conclusion
\begin{equation}\label{e27}
	\sum\limits_{i=1}^ne_i^k=\sum\limits_{i=1}^n(r_i^k-Mu_i^k),
\end{equation}
from which the first relation of \textit{Lemma 1} follows.

Using the same method, we can prove the second relation of \textit{Lemma 1}.

\textit{Lemma 2}: Under \textit{Assumption 2}, we have that $\Delta\bm{e}^{k+1}=W\bm{e}^k+(\bm{\delta}^{k+1}-\bm{\delta}^k)$, and $\Delta\bm{\mu}^{k+1}=W\Delta\bm{\mu}^k+\rho\Delta\bm{e}^{k+1}$.

\textit{Proof}: From the equation in line \textbf{6} in \textit{Algorithm \ref{a1}}, we can know that
\begin{equation}\label{e36}
	\bm{e}^{k+1}=W\bm{e}^k+(\bm{r}^{k+1}-\bm{r}^k)-(M\bm{u}^{k+1}-M\bm{u}^k).
\end{equation}

Combine (\ref{e36}) with \textit{Lemma 1}, we can get that
\begin{equation}\label{e37}
	\begin{aligned}
		\Delta\bm{e}^{k+1}&=W\bm{e}^k+(\bm{r}^{k+1}-\bm{r}^k)-(M\bm{u}^{k+1}-M\bm{u}^k)-\bar{\bm{e}}^{k+1}\\
		&=W(\bm{e}^k-\bar{\bm{e}}^k)+W\bar{\bm{e}}^k+\bm{\delta}^{k+1}-(\bm{r}^k-M\bm{u}^k).
	\end{aligned}
\end{equation}

Based on the properties of Kronecker product, we know that $W$ is doubly stochastic, thus 
\begin{equation}\label{e38}
	W\bar{\bm{e}}^k=\bar{\bm{e}}^k.
\end{equation}

Combine (\ref{e37}) with (\ref{e38}), then we can get to the first relation of \textit{Lemma 2}. Using the same method, we can prove the second relation of \textit{Lemma 2}.

Based on these two Lemmas, we can prove the following theorem.

\textit{Theorem 1}: Under \textit{Assumption 1} and \textit{Assumption 2}, the sequences $\left\lbrace \mu_i^k\right\rbrace $ and $\left\lbrace e_i^k\right\rbrace $ generated by \textit{Algorithm 1} all achieve consensus, i.e., 
$$\begin{cases}
	\lim\limits_{k\to\infty}\mu_i^k=\lim\limits_{k\to\infty}\bar{\mu}^k, \\
	\lim\limits_{k\to\infty}e_i^k=\lim\limits_{k\to\infty}\bar{e}^k=0,
\end{cases}\quad\forall i=1,2,\dots, n.$$

\textit{Proof}: Define
$$\begin{aligned}
	&g(r_i,u_i) \\
	&=(\phi_i^k)^T(r_i-Mu_i)+\frac{\rho}{2}\left\|(r_i-r_i^k)-M(u_i-u_i^k)+\sigma_i^k\right\|_2^2.
\end{aligned}$$
It follows from line \textbf{5} in \textit{Algorithm 1} that $r_i^{k+1}$ and $u_i^{k+1}$ are the optimal solution of $F_i(r_i,u_i)+g(r_i,u_i)$, which obey the following relation \cite{PPA}
\begin{equation}\label{e28}
	F_i(r_i,u_i)-F_i(r_i^{k+1},u_i^{k+1})+\left[\begin{matrix}
		r_i-r_i^{k+1} \\
		u_i-u_i^{k+1}
	\end{matrix}\right]\nabla g(r_i^{k+1},u_i^{k+1})\geqslant0,
\end{equation}
in which $\nabla g$ is the gradient of the function. Unfold (\ref{e28}), we can get
\begin{equation}\label{e29-pre1}
	\begin{aligned}
		&F_i(r_i,u_i)-F_i(r_i^{k+1})+\phi_i^k(r_i-Mu_i)-\phi_i^k(r_i^{k+1}-Mu_i^{k+1})\\
		&+\rho[r_i^{k+1}-r_i^k-M(u_i^{k+1}-u_i^k)+\sigma_i^k]^T[(r_i-r_i^{k+1})-\\
		&M(u_i-u_i^{k+1})\geqslant0,\quad\forall(r_i,u_i)\in\Omega_i.
	\end{aligned}
\end{equation}
Apply lines \textbf{6},\textbf{7} in \textit{Algorithm 1} into (\ref{e29-pre1}), we can get the following result
\begin{equation}\label{e29}
	\begin{aligned}
		&F_i(r_i^{k+1},u_i^{k+1})+(\mu_i^{k+1})^T(r_i^{k+1}-Mu_i^{k+1})\\
		\leqslant &F_i(r_i,u_i)+(\mu_i^{k+1})^T(r_i-Mu_i),\quad\forall(r_i,u_i)\in\Omega_i.
	\end{aligned}
\end{equation}

For the following derivation, we use $(\bm{r}_i^{*},\bm{u}^{*})$ to denote the optimal solution of the problem (\ref{e17}). Substitute $r_i=r_i^{*}$ and $u_i=u_i^{*}$ into (\ref{e29}), we can get
\begin{equation}\label{ee1}
	\begin{aligned}
		&F_i(r_i^{k+1},u_i^{k+1})+(\mu_i^{k+1})^T[(r_i^{k+1}-Mu_i^{k+1})-(r_i^{*}-Mu_i^{*})]\\
		\leqslant &F_i(r_i^{*},u_i^{*}).
	\end{aligned}
\end{equation}

Under \textit{Assumption 1}, based on the Saddle Point Theorem \cite{NP1995}, we can get that for $\forall (r_i,u_i)\in\Omega_i$,
\begin{equation}\label{ee2}
		L_i(r_i^{*},u_i^{*},\mu_i)\leqslant L_i(r_i^{*},u_i^{*},\mu_i^{*})\leqslant L_i(r_i,u_i,\mu_i^{*}),
\end{equation}
in which $L_i$ is the local Lagrange function of $\text{EH}_i$. It follows that
\begin{equation}
	L_i(r_i^{*},u_i^{*},\mu_i^{*})=F_i(r_i^{*},u_i^{*}), \quad\sum\limits_{i=1}^n(r_i^{*}-Mu_i^{*})=0.
\end{equation}
Let $\mu_i=\mu_i^{k+1}$ in (\ref{ee2}), summing (\ref{ee2}) up from $i=1$ to $i=n$, and noticing that
\begin{multline}\label{ee6}
		(\mu^{*})^T\sum\limits_{i=1}^n[(r_i-Mu_i)-(r_i^{*}-Mu_i^{*})]\\
		=(\bm{\mu}^{*})^T[(\bm{r}-\bm{Mu})-(\bm{r}^{*}-\bm{Mu}^{*})],
\end{multline}
in which $\bm{M}=\text{blkdiag}\{M\}_n$ is a block-diagonal matrix with $M$ placing only on the diagonal. Then we can get
\begin{equation}\label{e30}
	\begin{aligned}
		&F(\bm{r}^{k+1},\bm{u}^{k+1})+(\bm{\mu}^{k+1})^T[(\bm{r}^{k+1}-\bm{Mu}^{k+1})-(\bm{r}^{*}-\bm{Mu}^{*})]\\
		\leqslant &F(\bm{r},\bm{u})+(\bm{\mu}^{*})^T[(\bm{r}-\bm{Mu})-(\bm{r}^{*}-\bm{Mu}^{*})],
	\end{aligned}
\end{equation}

Let $\bm{r}=\bm{r}^{k+1}$, $\bm{u}=\bm{u}^{k+1}$ in (\ref{e30}), then we can get
\begin{equation}\label{e31}
	(\bm{\mu}^{k+1}-\bm{\mu}^{*})^T[(\bm{r}^{k+1}-\bm{Mu}^{k+1})-(\bm{r}^{*}-\bm{Mu}^{*})]\leqslant0.
\end{equation}

Since $\bm{\mu}^{k+1}-\bm{\mu}^{*}=(\bm{\mu}^{k+1}-\bar{\bm{\mu}}^{k+1})+(\bar{\bm{\mu}}^{k+1}-\bm{\mu}^{*})$, Eq.(\ref{e31}) can be rewritten as follows
\begin{equation}\label{e32}
	\begin{aligned}
		(\bm{\mu}^{k+1}-\bar{\bm{\mu}}^{k+1})^T[(\bm{r}^{k+1}-\bm{Mu}^{k+1})-(\bm{r}^{*}-\bm{Mu}^{*})]\\
		+(\bar{\bm{\mu}}^{k+1}-\bm{\mu}^{*})^T[(\bm{r}^{k+1}-\bm{Mu}^{k+1})-(\bm{r}^{*}-\bm{Mu}^{*})]\leqslant0.
	\end{aligned}
\end{equation}

For the first item of (\ref{e32}), we can get that
\begin{align}\label{e34}
		&(\bm{\mu}^{k+1}-\bar{\bm{\mu}}^{k+1})^T[(\bm{r}^{k+1}-\bm{Mu}^{k+1})-(\bm{r}^{*}-\bm{Mu}^{*})]\notag\\
		&=\left(\bm{\mu}^{k+1}-\bar{\bm{\mu}}^{k+1}\right)^T\left[\left(\bm{r}^{k+1}-\bm{Mu}^{k+1}\right)-\bar{\bm{e}}^{k+1}-\left(\bm{r^{*}}-\bm{Mu^{*}}\right)\right] \notag\\
		&={(\Delta\bm{\mu}^{k+1})}^T[\bm{\delta}^{k+1}-(\bm{r^{*}}-\bm{Mu^{*}})].
\end{align} 

The second item of (\ref{e32}) can be equally transformed to the following form
\begin{align}\label{e33}	
		&(\bar{\bm{\mu}}^{k+1}-\bm{\mu}^{*})^T[(\bm{r}^{k+1}-\bm{Mu}^{k+1})-(\bm{r}^{*}-\bm{Mu}^{*})]\notag\\
		&=(\bar{\mu}^{k+1}-\mu^{*})^T\sum\limits_{i=1}^n{(r_i^{k+1}-Mu_i^{k+1})}.
\end{align}
Applying \textit{Lemma 1} to (\ref{e33}), we can get
\begin{equation}\label{ee3}
	\begin{aligned}
		(\bar{\mu}^{k+1}-\mu^{*})^T\sum\limits_{i=1}^n{(r_i^{k+1}-Mu_i^{k+1})} \\
		=\frac{1}{\rho}\left(\bar{\bm{\mu}}^{k+1}-\bm{\mu}^{*}\right)^T\left(\bar{\bm{\mu}}^{k+1}-\bar{\bm{\mu}}^{k}\right).
	\end{aligned}	
\end{equation}

Combining (\ref{ee3}), (\ref{e34}) with (\ref{e32}), and noticing that
$$\begin{aligned}
	&2\left(\bar{\bm{\mu}}^{k+1}-\bm{\mu}^{*}\right)^T\left(\bar{\bm{\mu}}^{k+1}-\bar{\bm{\mu}}^{k}\right) \\
	&=\left\|\bar{\bm{\mu}}_{k+1}-\bm{\mu^{*}}\right\|_2^2+\left\|\bar{\bm{\mu}}^{k+1}-\bm{\bar{\mu}}^{k}\right\|_2^2-\left\|\bar{\bm{\mu}}^{k}-\bm{\mu}^{*}\right\|_2^2,
\end{aligned}$$
 we can obtain the following inequality
\begin{equation}\label{e35}
	\begin{aligned}
		\left\|\bar{\bm{\mu}}_{k+1}-\bm{\mu^{*}}\right\|_2^2+2\rho{(\Delta\bm{\mu}^{k+1})}^T[\bm{\delta}^{k+1}-(\bm{r^{*}}-\bm{Mu^{*}})]\\
		\leqslant \left\|\bar{\bm{\mu}}^{k}-\bm{\mu}^{*}\right\|_2^2-\left\|\bar{\bm{\mu}}^{k+1}-\bm{\bar{\mu}}^{k}\right\|_2^2.
	\end{aligned}	
\end{equation}

In the following, we will utilize the Lyapunov-based method to prove that consensus can be achieved asymptotically. Define $W_1=W-\mathcal{O}_{n}\otimes I_2$, and
$$\bm{z}^k=\left[\begin{matrix}
	\Delta\bm{\mu}^k & \rho\Delta\bm{e}^k
\end{matrix}\right]^T,\quad\quad\bm{c}^k=\rho[\bm{\delta}^{k+1}-(\bm{r^{*}}-\bm{Mu^{*}})],$$
$$\tilde{W}=\left[\begin{matrix}
	W_1 & W_1\\
	0 & W_1
\end{matrix}\right],\quad\quad\tilde{I}=\left[\begin{matrix}
I_{2n}\\
I_{2n}
\end{matrix}\right],\quad\quad H=\left[\begin{matrix}
I_{2n} & 0
\end{matrix}\right]_.$$

By \textit{Lemma 2} we can obtain the state-transition equation
\begin{equation}\label{e39}
	\bm{z}^{k+1}=\tilde{W}\bm{z}^k+\tilde{I}(\bm{c}^{k+1}-\bm{c}^k).
\end{equation}

Take a positive definite matrix $P\succ0$ and construct the Lyapunov function
$$V^k=\left\|\tilde{I}\bm{c}^{k}-\bm{z}^{k}\right\|_P^2+\left\|\bar{\bm{\mu}}^{k}-\bm{\mu^{*}}\right\|_2^2.$$

By Lyapunov stability theory, if there exists a $P\succ0$ such that $V^{k+1}-V^k\leqslant 0$, i.e.
\begin{equation}\label{ee5}
	\begin{aligned}
		\left\|\tilde{I}\bm{c}^{k+1}-\bm{z}^{k+1}\right\|_P^2+\left\|\bar{\bm{\mu}}^{k+1}-\bm{\mu^{*}}\right\|_2^2 \\
		\leqslant\left\|\tilde{I}\bm{c}^{k}-\bm{z}^{k}\right\|_P^2+\left\|\bar{\bm{\mu}}^{k}-\bm{\mu^{*}}\right\|_2^2, \\
	\end{aligned}	
\end{equation}
then the system (\ref{e39}) is Lyapunov stable.
 Noticing that
\begin{equation*}\label{e40}
	\begin{aligned}
		&\left\|\tilde{I}\bm{c}^{k+1}-\bm{z}^{k+1}\right\|_P^2=\left\|\tilde{I}\bm{c}^{k}-\bm{z}^{k}+(I_{4n}-\tilde{W})\bm{z}^k\right\|_P^2\\
		&=2(\bm{c}^k)^T\tilde{I}^TP(I_n-\tilde{W})\bm{z}^k+\left\|\tilde{I}\bm{c}^{k}-\bm{z}^{k}\right\|_P^2-\left\|z^k\right\|_{P-\tilde{W}^TP\tilde{W}.}^2
	\end{aligned}
\end{equation*}

Add the above equation to (\ref{e35}), we can get
\begin{equation}\label{e41}
	\begin{aligned}
		&\left\|\bar{\bm{\mu}}^{k+1}-\bm{\mu^{*}}\right\|_2^2+2(\bm{c}^{k+1})^TH{\bm{z}}^{k+1}+\left\|\tilde{I}\bm{c}^{k+1}-\bm{z}^{k+1}\right\|_P^2\\
		\leqslant&\left\|\bar{\bm{\mu}}^{k}-\bm{\mu^{*}}\right\|_2^2+2(\bm{c}^{k+1})^T\tilde{I}^TP(I_{4n}-\tilde{W}){\bm{z}}^{k+1}+\left\|\tilde{I}\bm{c}^{k}-\bm{z}^{k}\right\|_P^2\\
		&-\left\|\bar{\bm{\mu}}^{k+1}-\bm{\bar{\mu}^{k}}\right\|_2^2-\left\|\bm{z}^k\right\|_{P-\tilde{W}^TP\tilde{W}.}^2
	\end{aligned}
\end{equation}

Compare (\ref{ee5}) with (\ref{e41}), we find that if there exists a $P$ satisfying the following conditions
\begin{subequations}\label{e42}
\begin{align}
	&\tilde{I}^TP(I_{4n}-\tilde{W})=H,\\
	&P-\tilde{W}^TP\tilde{W}\succ0,\\
	&P\succ0,
\end{align}
\end{subequations}
then $V^{k+1}\leqslant V^k$, $\forall k$. This means that the sequence $\left\lbrace V^k\right\rbrace $ is monotonically decreasing. Since $V^k\geqslant0$, then $\left\|V^{k}-V^{k+1}\right\|\to0$ as $k\to\infty$.
Based on the work in \cite{2020Tracking} and \cite{2004Fast}, we know that $W$ is symmetric and doubly stochastic, $\Gamma(W_1)<1$, and $(I_{2n}-W_1)$ is reversible. Therefore, we can apply the work in \cite{2018A} and get the following $P^{*}$ which meets the condition (\ref{e42})
$$\left[\begin{matrix}
	2I_{2n} & (I_{2n}-W_1)^{-1}-2I_{2n}\\
	(I_n-W_1)^{-1}-2I_{2n} & (I_n-W_1)^{-2}-2(I_{2n}-W_1)^{-1}+2I_{2n}
\end{matrix}\right]_.$$

Let $P=P^{*}$, then (\ref{e41}) can be rewritten as
\begin{equation}\label{e43}
	0\leqslant\left\|\bar{\bm{\mu}}^{k+1}-\bm{\bar{\mu}^{k}}\right\|_2^2+\left\|\bm{z}^k\right\|_{P^{*}-\tilde{W}^TP^{*}\tilde{W}}^2\leqslant\left\|V^{k}-V^{k+1}\right\|.
\end{equation}
  
Since $\left\|V^{k+1}-V^k\right\|\to0$, (\ref{e43}) equals to $\lim\limits_{k\to\infty}\left\|\bm{\bar{\mu}}^k-\bar{\bm{\mu}}^{k+1}\right\|=0$ and $\lim\limits_{k\to\infty}\left\|\bm{z}^k\right\|=0$. This means
\begin{equation}\label{e44}
		\lim\limits_{k\to\infty}\mu_i^k=\lim\limits_{k\to\infty}\bar{\mu}^k,\quad	\lim\limits_{k\to\infty}e_i^k=\lim\limits_{k\to\infty}\bar{e}^k,
\quad \forall i=1,2,\dots,n.
\end{equation}

According to (\ref{e44}) and \textit{Lemma 1}, we can get that 
\begin{equation}\label{e45}
	\lim\limits_{k\to\infty}\left\|\bar{e}^k\right\|=\lim\limits_{k\to\infty}\left\|\frac{1}{\rho}({\bar{\mu}}^{k-1}-\bar{{\mu}}^{k})\right\|=0.
\end{equation}

This completes the proof.

\textit{Theorem 2}: Under \textit{Assumption 1} and \textit{Assumption 2}, the sequences $\left\lbrace r^k\right\rbrace $ and $\left\lbrace u^k\right\rbrace $ generated by \textit{Algorithm \ref{a1}} converge to the optimal value of the global welfare $F^{*}$, i.e.,  $$\lim\limits_{k\to\infty}F(\bm{r}^k,\bm{u}^k)=F^{*}.$$

\textit{Proof}: It follows from (\ref{ee2}) and (\ref{ee6}) that
\begin{align}\label{e46}
			&F(\bm{r}^{k+1},\bm{u}^{k+1})+(\bm{\mu}^{k+1})^T[(\bm{r}^{k+1}-\bm{Mu}^{k+1})-(\bm{r}^{*}-\bm{Mu}^{*})] \notag\\
			&\leqslant F(\bm{r}^{*},\bm{u}^{*}).
\end{align}
Rewrite $\bm{\mu}^{k+1}$ in (\ref{e46}) as $\bar{\bm{\mu}}^{k+1}+\Delta\bm{\mu}^{k+1}$, and noticing that
\begin{align}\label{e47}
		&(\bar{\bm{\mu}}^{k+1})^T[(\bm{r}^{k+1}-\bm{Mu}^{k+1})-(\bm{r}^{*}-\bm{Mu}^{*})] \notag\\
		&=(\bar{{\mu}}^{k+1})^T\sum\limits_{i=1}^n[(r_i^{k+1}-Mu_i^{k+1})-(r_i^{*}-Mu_i^{*})] \notag\\
		&=(\bar{{\mu}}^{k+1})^T\sum\limits_{i=1}^n[(r_i^{k+1}-Mu_i^{k+1})].
\end{align}
Apply \textit{Lemma 1} to (\ref{e47}), we can get that
\begin{equation}\label{e48}
		(\bar{\bm{\mu}}^{k+1})^T[(\bm{r}^{k+1}-\bm{Mu}^{k+1})-(\bm{r}^{*}-\bm{Mu}^{*})]=n(\bar{{\mu}}^{k+1})^T\bar{e}^{k+1}.
\end{equation}
Combining (\ref{e48}) with (\ref{e46}) yields
\begin{equation}\label{e49}
	\begin{aligned}
		&F(\bm{r}^{k+1},\bm{u}^{k+1})+(\bm{\mu}^{k+1})^T[(\bm{r}^{k+1}-\bm{Mu}^{k+1})-(\bm{r}^{*}-\bm{Mu}^{*})] \\
		=&F(\bm{r}^{k+1},\bm{u}^{k+1})+n(\bar{{\mu}}^{k+1})^T\bar{e}^{k+1} \\
		&+(\Delta\bm{\mu}^{k+1})^T[(\bm{r}^{k+1}-\bm{Mu}^{k+1})-(\bm{r}^{*}-\bm{Mu}^{*})]\\
		\leqslant &F(\bm{r}^{*},\bm{u}^{*}).
	\end{aligned}
\end{equation}

From \textit{Theorem 1} we know that $\lim\limits_{k\to\infty}\left\|\Delta\bm{\mu}^{k+1}\right\|=0$ and $\lim\limits_{k\to\infty}\left\|\bar{e}^{k+1}\right\|=0$. Taking the limits of both sides of (\ref{e49}) yields
\begin{equation}\label{e50}
	\lim\limits_{k\to\infty}F(\bm{r}^{k+1},\bm{u}^{k+1})\leqslant F(\bm{r}^{*},\bm{u}^{*}).
\end{equation}

From line \textbf{5} in \textit{Algorithm 1} we know that for $\forall k=1,2,\dots$, we all have $({r_i}^{k+1},{u_i}^{k+1})\in\Omega_i$. That means
\begin{equation}\label{e51}
	\lim\limits_{k\to\infty}F(\bm{r}^{k+1},\bm{u}^{k+1})\geqslant F(\bm{r}^{*},\bm{u}^{*}).
\end{equation}

Combine (\ref{e50}) and (\ref{e51}), we can get to the conclusion of \textit{Theorem 2}.


\begin{table*}[htbp]
	\caption{PARAMETERS OF LOCAL WELFARE FUNCTIONS.}
	\label{t1}
	\centering
	\begin{threeparttable}
		\begin{tabular}{||c||p{0.5cm}|p{0.5cm}|p{0.5cm}|p{0.5cm}|p{0.5cm}|p{0.5cm}|p{0.5cm}|p{0.5cm}|p{0.5cm}|p{0.5cm}|p{0.5cm}|p{0.5cm}|p{0.5cm}|p{0.5cm}||}
			\hline
			\hline
			Node & N1 & N2 & N3 & N4 & N5 & N6 & N7 & N8 & N9 & N10 & N11 & N12 & N13 & N14\\
			\hline
			$a_{1i}^e$(mu/pu$^2$) & 0.11 & 0.05 & 0.08 & 0.03 & 0.06 & 0.07 & 0.04 & 0.12 & 0.11 & 0.06 & 0.09 & 0.05 & 0.07 & 0.08 \\
			\hline
			$a_{2i}^e$(mu/pu) & 12.0 & 13.5 & 11.5 & 12.5 & 11.7 & 11.9 & 12.6 & 12.8 & 11.6 & 13.3 & 13.2 & 13.0 & 12.7 & 12.1 \\
			\hline
			$a_{1i}^g$(mu/pu$^2$) & 0.033 & 0.042 & 0.033 & 0.021 & 0.034 & 0.025 & 0.028 & 0.036 & 0.030 & 0.029 & 0.023 & 0.027 & 0.026 & 0.031 \\
			\hline
			$a_{2i}^g$(mu/pu) & 5.6 & 5.0 & 5.5 & 6.6 & 5.7 & 5.5 & 5.3 & 6.1 & 6.4 & 6.0 & 5.8 & 5.9 & 5.1 & 5.2 \\
			\hline
			$\gamma_{1i}^e$(mu/pu$^2$) & 0.13 & 0.14 & 0.11 & 0.09 & 0.15 & 0.16 & 0.10 & 0.12 & 0.13 & 0.08 & 0.11 & 0.07 & 0.11 & 0.10\\
			\hline
			$\gamma_{2i}^e$(mu/pu) & 7.2 & 7.3 & 8.5 & 7.4 & 7.7 & 8.1 & 8.2 & 7.9 & 8.0 & 7.5 & 7.6 & 7.4 & 7.8 & 8.3\\
			\hline
			$\gamma_{1i}^g$(mu/pu$^2$) & 0.023 & 0.024 & 0.030 & 0.028 & 0.015 & 0.017 & 0.020 & 0.022 & 0.016 & 0.018 & 0.021 & 0.017 & 0.026 & 0.010\\
			\hline
			$\gamma_{2i}^g$(mu/pu) & 3.4 & 3.3 & 4.5 & 3.7 & 3.8 & 4.1 & 3.2 & 3.9 & 4.3 & 3.6 & 3.8 & 4.0 & 3.9 & 3.9\\
			\hline
			constant item(pu)\tnote{1} & 0.57 & 0.33 & 0.50 & 0.58 & 0.21 & 0.24 & 0.72 & 0.15 & 0.78 & 0.22 & 0.40 & 0.56 & 0.42 & 0.53 \\
			\hline
			$[\underline{\xi_i^e},\overline{\xi_i^e}]$(10pu)\tnote{2} & [2,9] & [4,15] & [1,10] & [2,14] & [2,14] & [3,15] & [3,15] & [4,16] & [2,15] & [0,9] & [5,13] & [2,15] & [2,15] & [3,16]\\
			\hline
			$[\underline{\xi_i^g},\overline{\xi_i^g}]$(10pu) & [3,10] & [2,16] & [1,12] & [2,14] & [3,16] & [4,17] & [3,15] & [3,16] & [2,14] & [0,10] & [5,15] & [1,14] & [2,16] & [3,16]\\
			\hline
			\multicolumn{15}{||c||}{For all the nodes, $\eta_i^{ee}=0.9$, $\eta_i^{ce}=0.7$, $\eta_i^{ch}=0.5$, $\eta_i^{gh}=0.4$, $\zeta^e=1.1$(\$/pu), $\zeta^g=0.6$(\$/pu)}\\
			\hline
			\hline
		\end{tabular}
		\begin{tablenotes}    
			\footnotesize
			\item[1] Constant item $=a_{3i}^e+a_{3i}^g+\gamma_{3i}^e+\gamma_{3i}^g$. We merge all the constant items since they have no influence on our algorithm.
			\item[2] $\xi=r,s,d$. We set the same upper-bound and lower-bound for $r_i$, $s_i$ and $d_i$ in the same node.      
		\end{tablenotes}
	\end{threeparttable}
\end{table*}

\section{CASE STUDY}\label{s5}

In this section, we present a case study to examine the facility and effectiveness of our algorithm.

\begin{figure}[t]	
	\centering
	\includegraphics[width=0.4\textwidth]{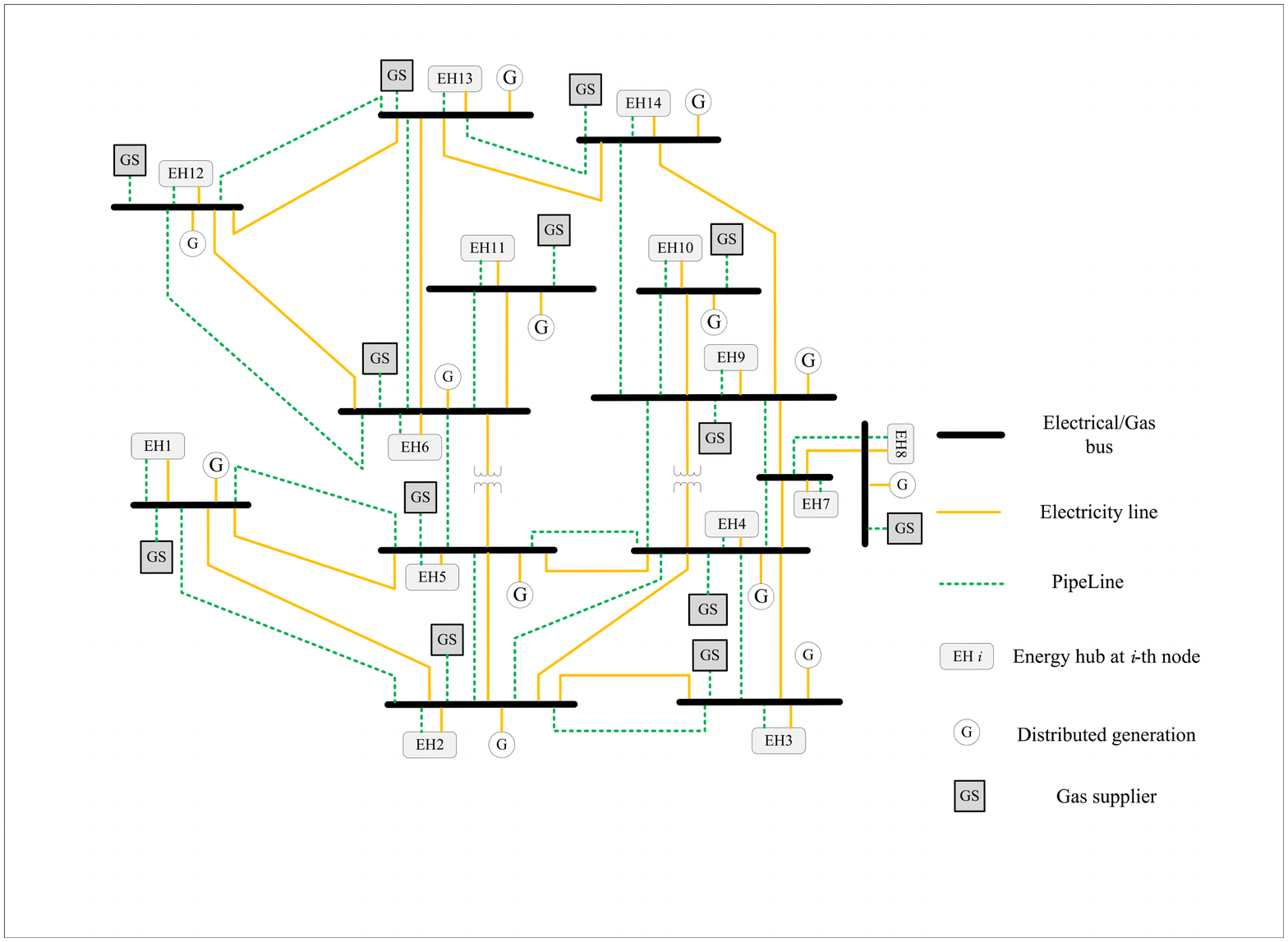}
	\caption{Structure of Simulation System.}
	\label{fig-3}
\end{figure}

We choose an MES  based on the modified IEEE 14-bus network. Each bus contains an EH and the system contains a pipeline network and an electricity network which connect all the buses. Fig.(\ref{fig-3}) shows the structure of the system. 

For the simulations we choose the quadratic functions as follows
$$Ce_i(r_i^{e})=a_{1i}^{e}(r_i^{e})^2+a_{2i}^e r_i^{e}+a_{3i}^{e},\quad a_{1i}^e,a_{2i}^e,a_{3i}^e>0,$$
$$U_i(d_i^{e})=-\gamma_{1i}^{e}(d_i^{e})^2+\gamma_{2i}^e d_i^{e}+\gamma_{3i}^{e},\quad \gamma_{1i}^{e},\gamma_{2i}^{e},\gamma_{3i}^{e}>0,$$
$$Cg_i(r_i^{g})=a_{1i}^{g}(r_i^{g})^2+a_{2i}^e r_i^{g}+a_{3i}^{g},\quad a_{1i}^g,a_{2i}^g,a_{3i}^g>0,$$
$$U_i(d_i^{g})=-\gamma_{1i}^{g}(d_i^{g})^2+\gamma_{2i}^g d_i^{g}+\gamma_{3i}^{g},\quad \gamma_{1i}^{g},\gamma_{2i}^{g},\gamma_{3i}^{g}>0.$$

The specific parameters can be found in Table \ref{t1}.

Set $\epsilon=0.05$ and $N=300$ and apply the proposed algorithm to the above SWOP. Fig.4(a) and Fig.4(b) show how $\frac{\left\|r_i^k-r_i^{*}\right\|}{\left\|r_i^{*}\right\|}$ and $\frac{\left\|d_i^k-d_i^{*}\right\|}{\left\|d_i^{*}\right\|}$ change with iteration $k$. As a comparison, we use the Lagrange method in the centralized form directly to get the optimal solution, which is not difficult in this case. It can be seen that optimization variables $r_i$ on supply side and $d_i$ on demand side of $\text{EH}_i$ could converge to the optimal solution.

\begin{figure}\label{fig-4}
	\centering
	\subfigure[Convergence of $r_i^k$]{
		\begin{minipage}[t]{1\linewidth}
			\centering
			\includegraphics[width=2in]{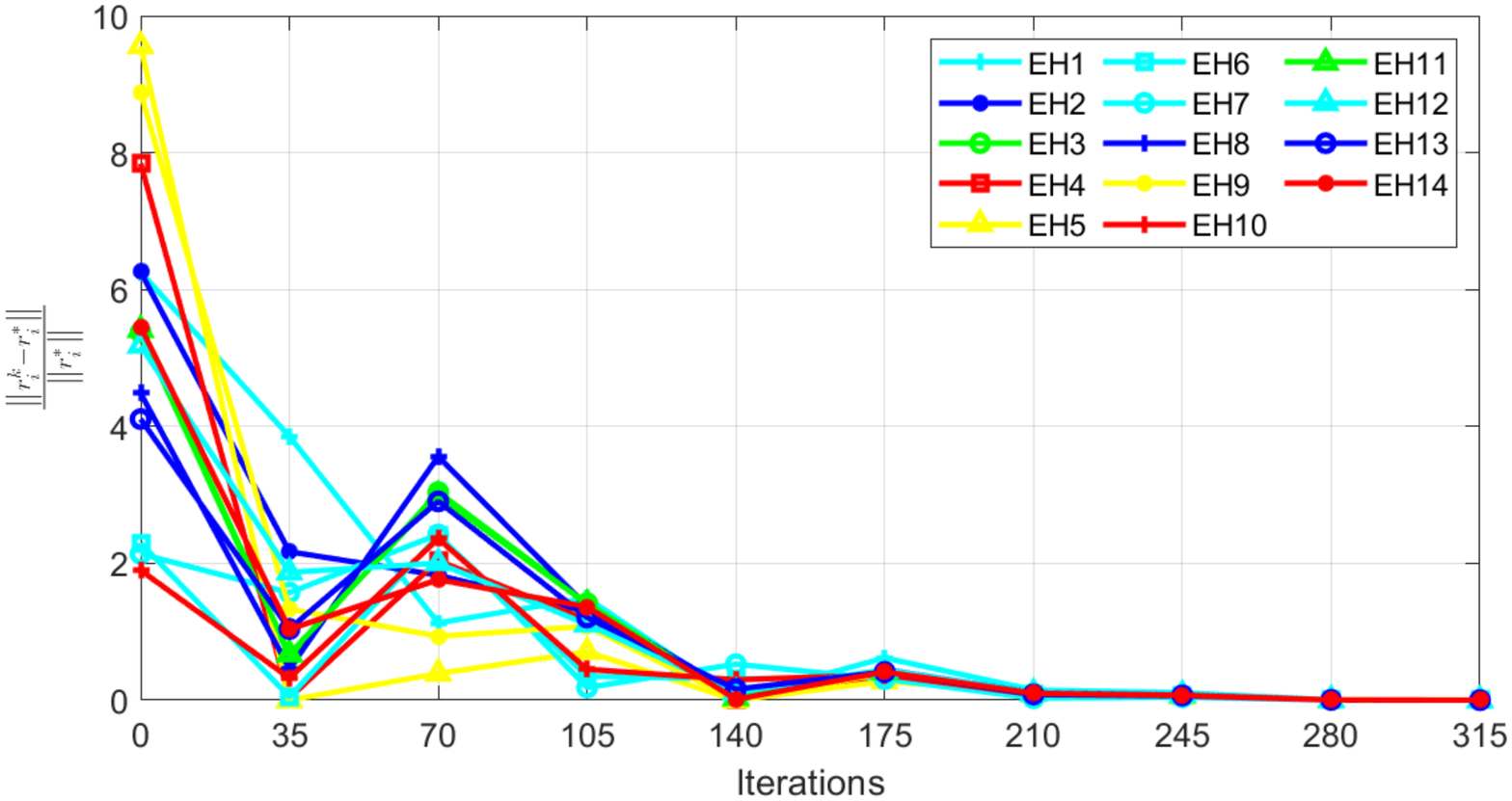}
		\end{minipage}%
	}%

	\subfigure[Convergence of $d_i^k$]{
		\begin{minipage}[t]{1\linewidth}
			\centering
			\includegraphics[width=2in]{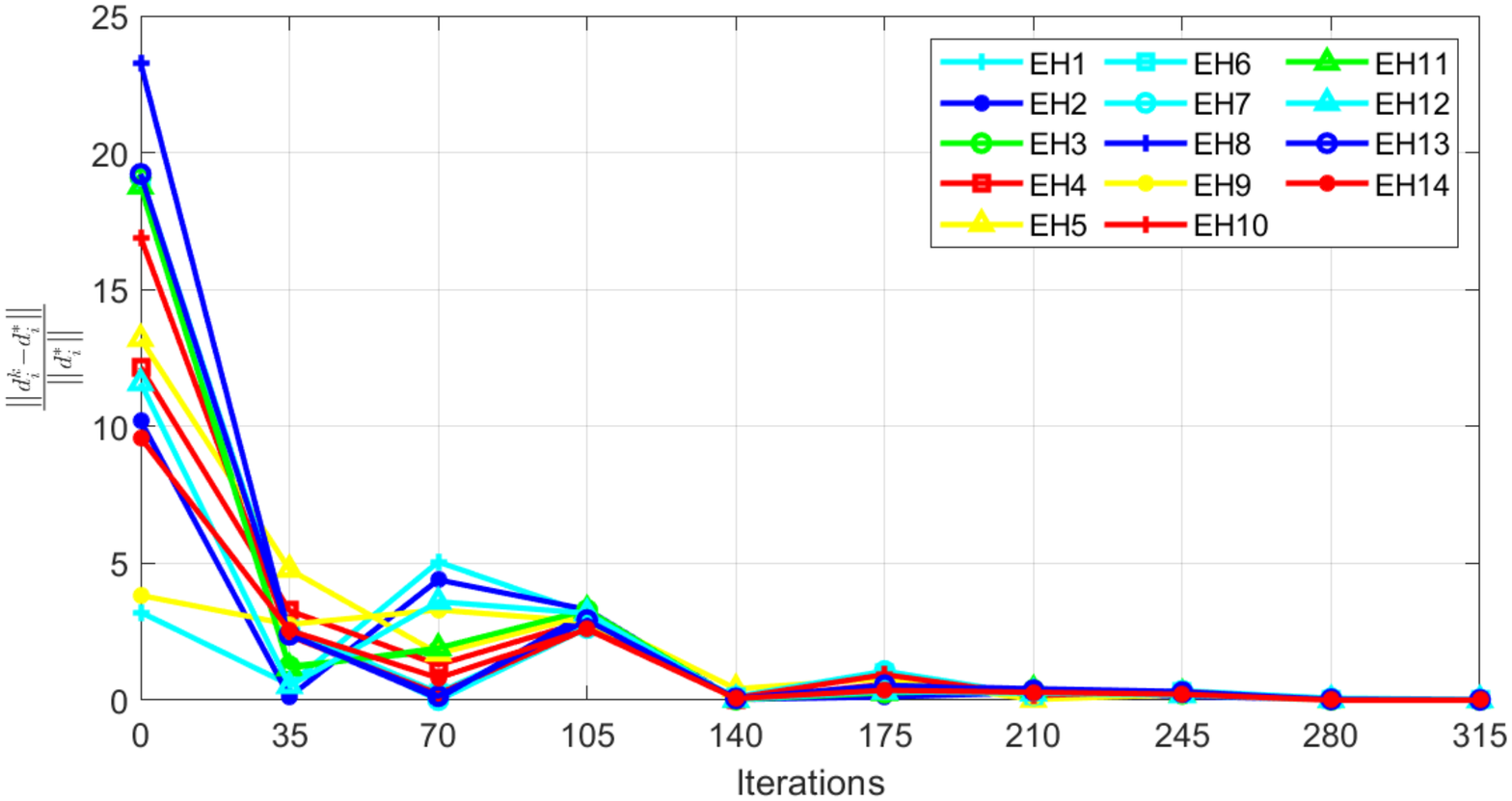}
		\end{minipage}%
	}%
	\centering
	\caption{The historical evolution of $\frac{\left\|r_i^k-r_i^{*}\right\|}{\left\|r_i^{*}\right\|}$ and $\frac{\left\|d_i^k-d_i^{*}\right\|}{\left\|d_i^{*}\right\|}$ with $k$.}
\end{figure}

\begin{figure}[t]\label{fig-5}
	\centering
	\subfigure[Electricity]{
		\begin{minipage}[t]{0.5\linewidth}
			\centering
			\includegraphics[width=1.5in]{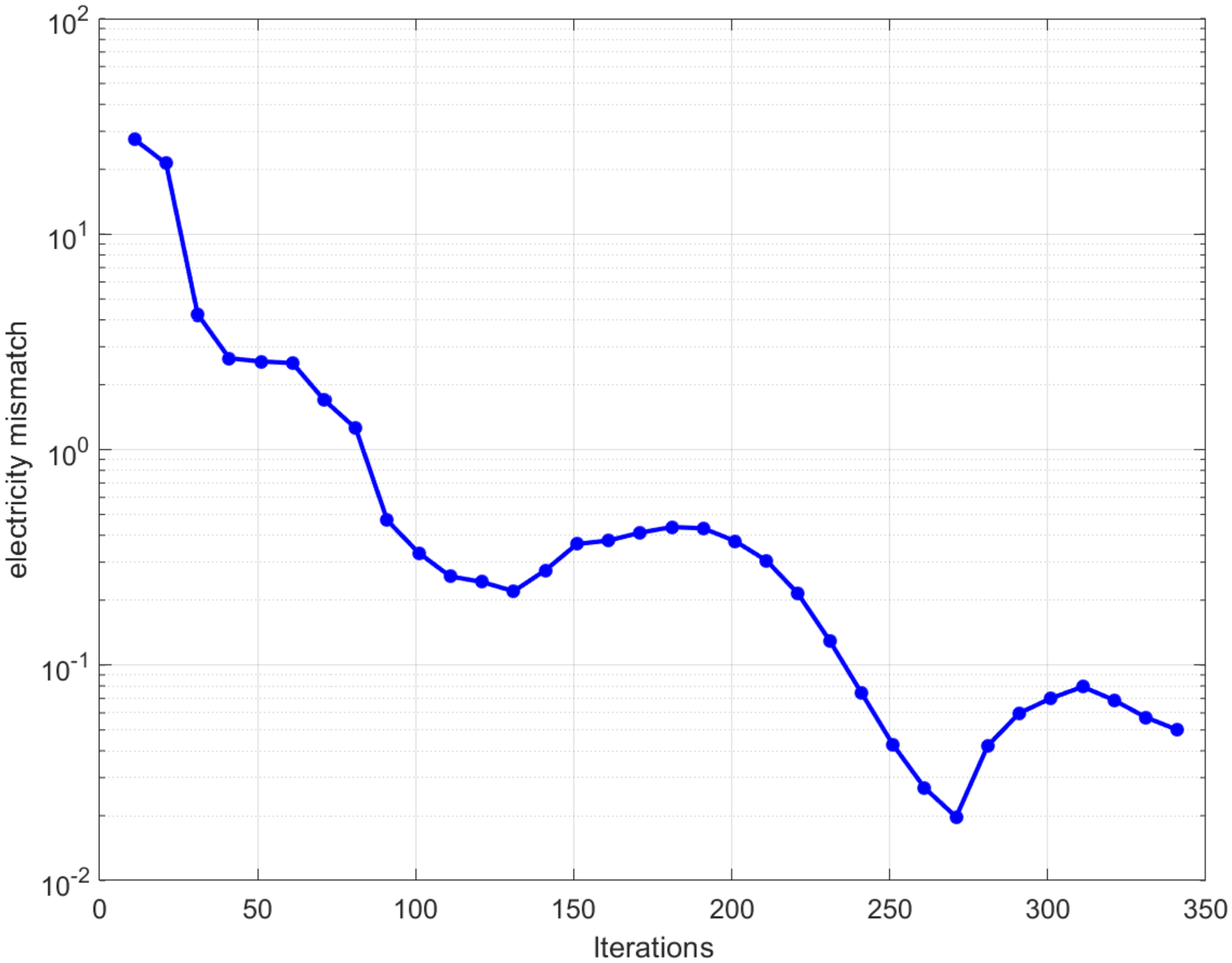}
		\end{minipage}%
	}%
	\subfigure[Gas]{
		\begin{minipage}[t]{0.5\linewidth}
			\centering
			\includegraphics[width=1.5in]{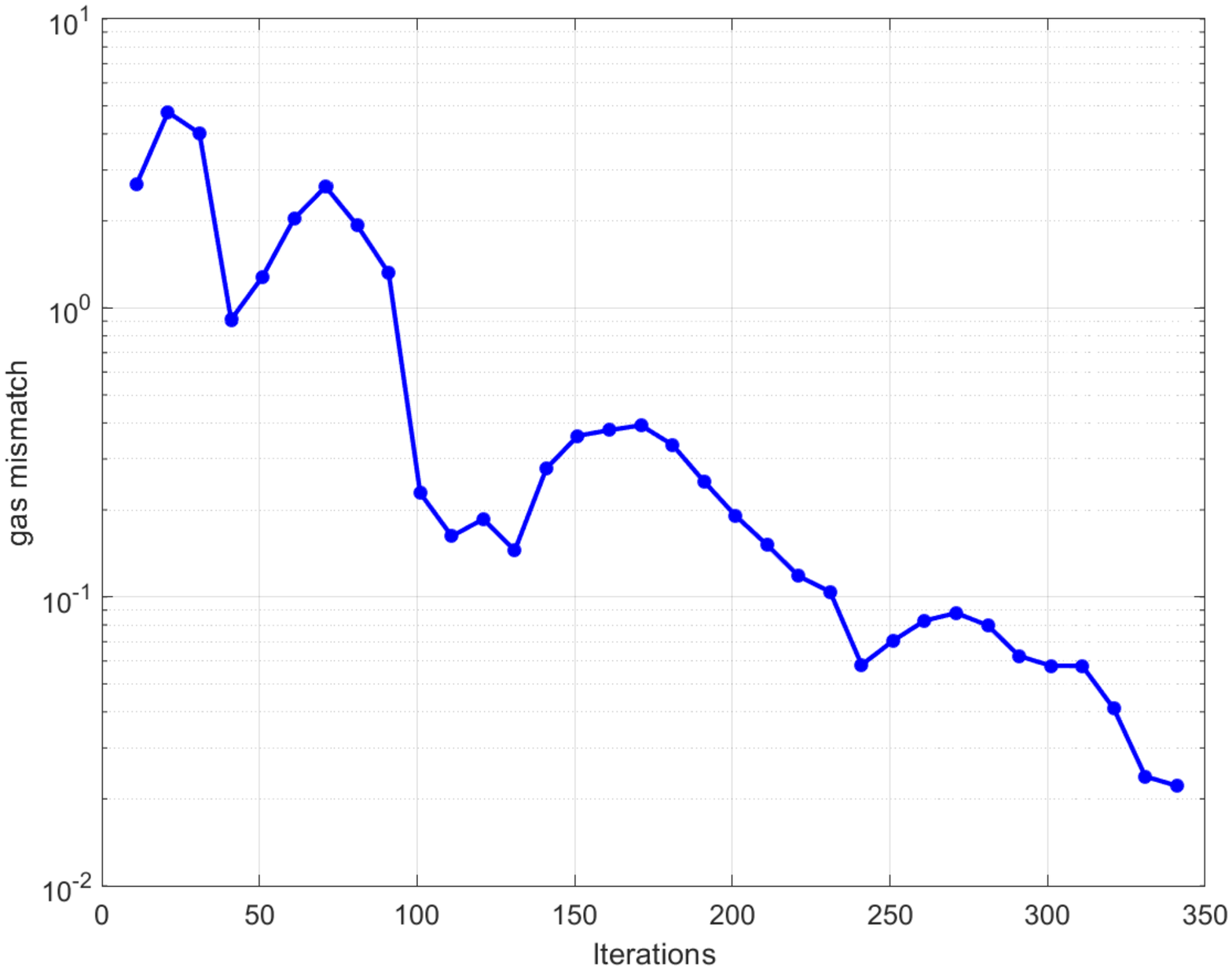}
		\end{minipage}%
	}%
	\centering
	\caption{Historical evolution of the total electricity and gas mismatches between
		energy suppliers and demand side.}
\end{figure}

 Next, we examine how the energy mismatches change through iterations. We define the energy mismatches of electricity and gas as $\sum_{i=1}^n[(r_i^e)^k-(s_i^e)^k]$ and $\sum_{i=1}^n[(r_i^g)^k-(s_i^g)^k]$, which reveals the energy mismatches between energy suppliers and energy users in the MES. Fig.5 shows that the energy mismatches of electricity and gas converge to $0$ at an acceptable rate.

\begin{figure}[t]
	\centering
	\includegraphics[width=0.35\textwidth]{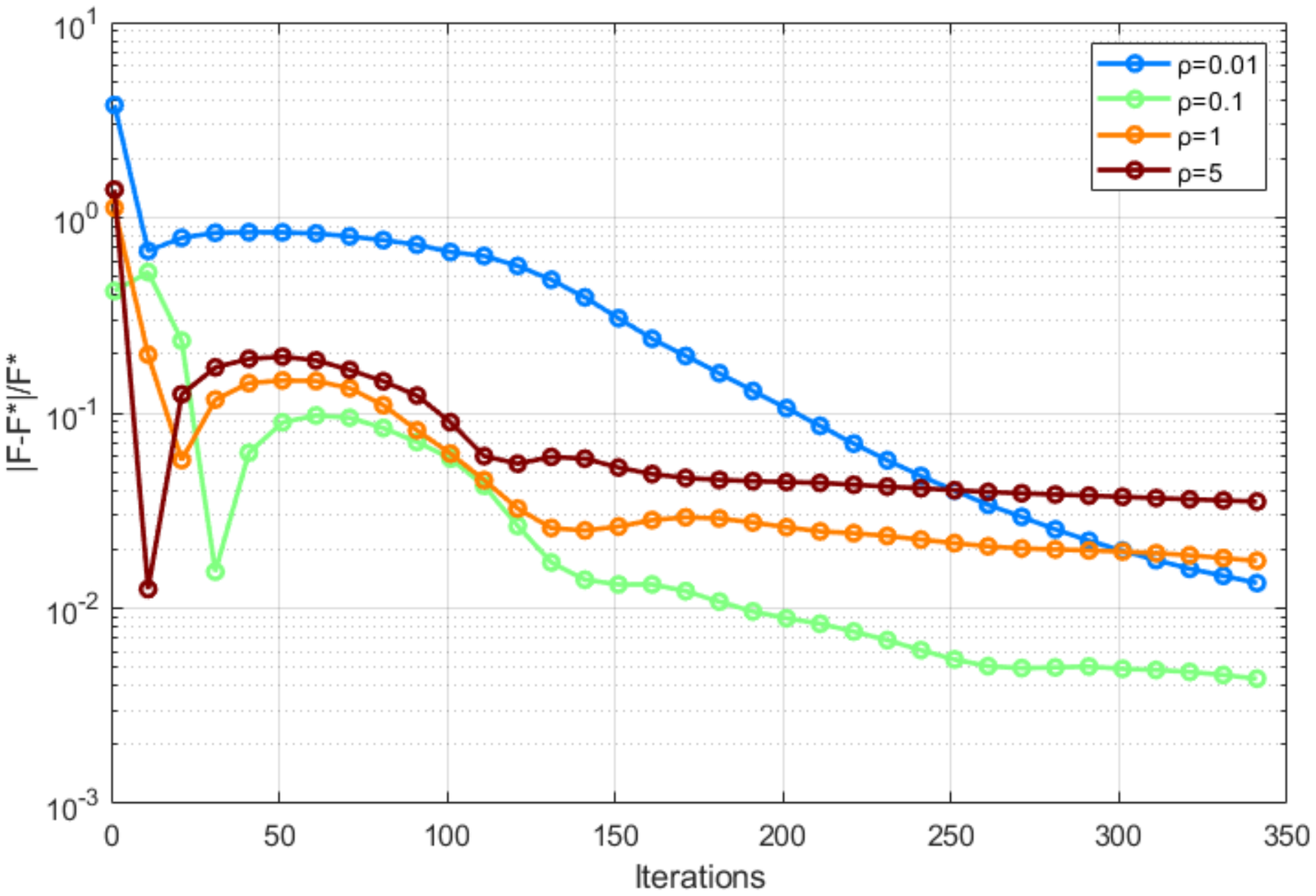}
	\caption{ Convergence to the optimal solution $F^{*}$ with different $\rho$.}
	\label{fig-6}
\end{figure}

Finally, we examine the effect of the penalty factor $\rho$ on the convergence of our algorithm. We choose $\rho=0.01,0.1,1,5$. Fig.\ref{fig-6} depicts the simulation result of $(F^k-F^{*})/F^{*}$ with iteration $k$. It is observed that the algorithm converges for all the chosen penalty factors. Moreover, increasing $\rho$ will accelerate the convergence but if $\rho$ is too large, it tends to diverge. It indicates that the penalty factor $\rho$ should be set properly and cannot be too large or too small. How to determine the proper $\rho$ needs further investigation.

\section{Conclusion}\label{s6}
In this paper we have investigated the coordination problem between ED and DR in MESs. We have formulated the problem as a SWOP, handled its nonlinearity and nonconvexity by using dimension augmentation techniques, and transformed it to a convex optimization problem. A decentralized algorithm based on parallel ADMM and dynamic average tracking has been proposed, and its convergence has been proved by using variational inequality and Lyapunov-based techniques. The effectiveness has been examined by a case study on the modified IEEE 14-bus network.
	\bibliographystyle{unsrt}
	\bibliography{paper}	
\end{document}